# Mid-infrared reconfiguration of population flow in lanthanide nanocrystals

Xinyang Yu[1,2], Yin Huang[1], Karin Yamamura[1,2], Chenyi Wang[3], Lei Ding[4], Mehran Kianinia[1,2], Yang Yu[5], Jiyun Kim[1,2], Baolei Liu[6], Xiaoxue Xu[6], Otto Cranwell Schaeper[1,2], Yue Bian[5], Lan Fu[5], Guochen Bao[1], Qian Peter Su[6], Fan Wang[3], Igor Aharonovich[1,2]*, Chaohao Chen[1,2,6]*

[1] School of Mathematical and Physical Sciences, Faculty of Science, The University of Technology Sydney, Ultimo, NSW 2007, Australia

[2] Australian Research Council Centre of Excellence for Transformative Meta-Optical Systems, The University of Technology Sydney, Ultimo, NSW 2007, Australia

[3] Science and Technology Center for Quantum Biology, National Institute of Extremely-Weak Magnetic Field Infrastructure, Hangzhou, China

[4] Centre for Atomaterials and Nanomanufacturing, School of Science, RMIT University, Melbourne, Victoria, Australia

[5] Australian Research Council Centre of Excellence for Transformative Meta-Optical Systems, Department of Electronic Materials Engineering, Research School of Physics, The Australian National University, Canberra ACT 2600, Australia

[6] School of Electrical, Mechanical and Biomedical Engineering, Faculty of Engineering and Information Technology, The University of Technology Sydney, Sydney, NSW 2007, Australia

* Correspondence to: Igor.Aharonovich@uts.edu.au; Chaohao.Chen@uts.edu.au

## Abstract

Converting mid-infrared (MIR) radiation to visible or near-infrared wavelengths is essential for imaging and sensing, yet achieving sensitive, low-power and scalable detection remains challenging. Here we show that MIR irradiation reconfigures population flow in lanthanide nanocrystals through state-dependent changes in excited-state relaxation kinetics. Time-resolved measurements reveal markedly different changes in the apparent decay rates of individual manifolds. The resulting change in kinetic balance produces anticorrelated population redistribution that cannot be reproduced by uniform heating alone. We exploit this kinetic redistribution to achieve a linear ratiometric response to MIR power across 6.8–8.6 µm. The ratiometric signal remains largely independent of the near-infrared pump power, enabling operation at excitation powers down to 10 µW. Using standard silicon photodetectors, we demonstrate room-temperature MIR detection with a detection limit approaching 4 nW µm$^{-2}$, together with reversible optical responses down to the single-nanocrystal level. Our results establish lanthanide nanocrystals as optically reconfigurable nanoscale energy-flow networks, in which low-energy MIR radiation reconfigures excited-state population flow for ratiometric MIR transduction.

## Introduction

Mid-infrared (MIR) detection is essential for many practical technologies because MIR photons encode rich molecular vibrational and thermal information.[1,2] Applications including chemical sensing[3] and environmental monitoring[4] and biomedical diagnostics[5] rely on sensitive detection of MIR radiation. However, the development of MIR detectors has lagged their visible and near-infrared counterparts because the low photon energies in the MIR are poorly matched to conventional semiconductor bandgaps.[6,7] As a result, high-performance MIR photodetectors often require cryogenic cooling or complex device architectures involving sophisticated nanofabrication.

Spectral conversion provides an attractive alternative by converting MIR radiation into visible or near-infrared photons that can be efficiently detected using mature silicon-based photodetectors.[8-11] However, most MIR-to-visible conversion approaches rely on nonlinear optical processes that require phase matching in bulk crystals or strong electromagnetic-field confinement in resonant nanostructures.[12-14] These requirements can impose constraints on device geometry, spectral bandwidth and large-area implementation.

Lanthanide nanocrystals provide a complementary route to MIR spectral transduction because their shielded 4f electronic manifolds support spectrally stable and narrowband emission, while Stark splitting creates closely spaced electronic sublevels that can couple to low-energy optical and vibrational excitations.[15-21] Excitation, energy transfer and phonon-assisted relaxation collectively define the population flow through these interconnected excited-state manifolds, determining how absorbed energy is redistributed among competing emissive states. In thermally coupled manifolds, population exchange can approach quasi-thermal distributions governed strongly by lattice temperature.[22-26] Recent studies have demonstrated MIR-to-optical transduction in lanthanide nanocrystals, particularly in $Nd^{3+}$- and $Nd^{3+}/Yb^{3+}$-based systems, through excited-state excitation[27,28], energy-transfer processes[29] or ligand-mediated photothermal effects.[30,31] These studies establish the potential of lanthanide nanotransducers for room-temperature MIR sensing. Whether MIR radiation can reconfigure population flow within coupled excited-state manifolds through state-dependent changes in their relaxation kinetics, distinct from conventional thermal responses, remains unclear.

Here we show that MIR irradiation reconfigures population flow within the coupled excited-state network of $Yb^{3+}/Er^{3+}$ nanocrystals, producing anticorrelated redistribution between thermally coupled $Er^{3+}$ emission channels. Time-resolved measurements reveal strongly state-dependent changes in the apparent decay rates, linking the population redistribution to changes in excited-state kinetics. This redistribution remains largely decoupled from the near-infrared excitation power, enabling a linear ratiometric response to MIR power across 6.8–8.6 μm and operation at excitation powers down to 10 μW. Using silicon-compatible visible readout, we achieve a MIR detection limit of approximately 4 nW $\mu m^{-2}$ and reversible responses down to the single-nanocrystal level. These results establish MIR-driven reconfiguration of population flow as a mechanism for low-power ratiometric transduction in lanthanide nanocrystals.

## Results

### MIR reconfigures excited-state population flow

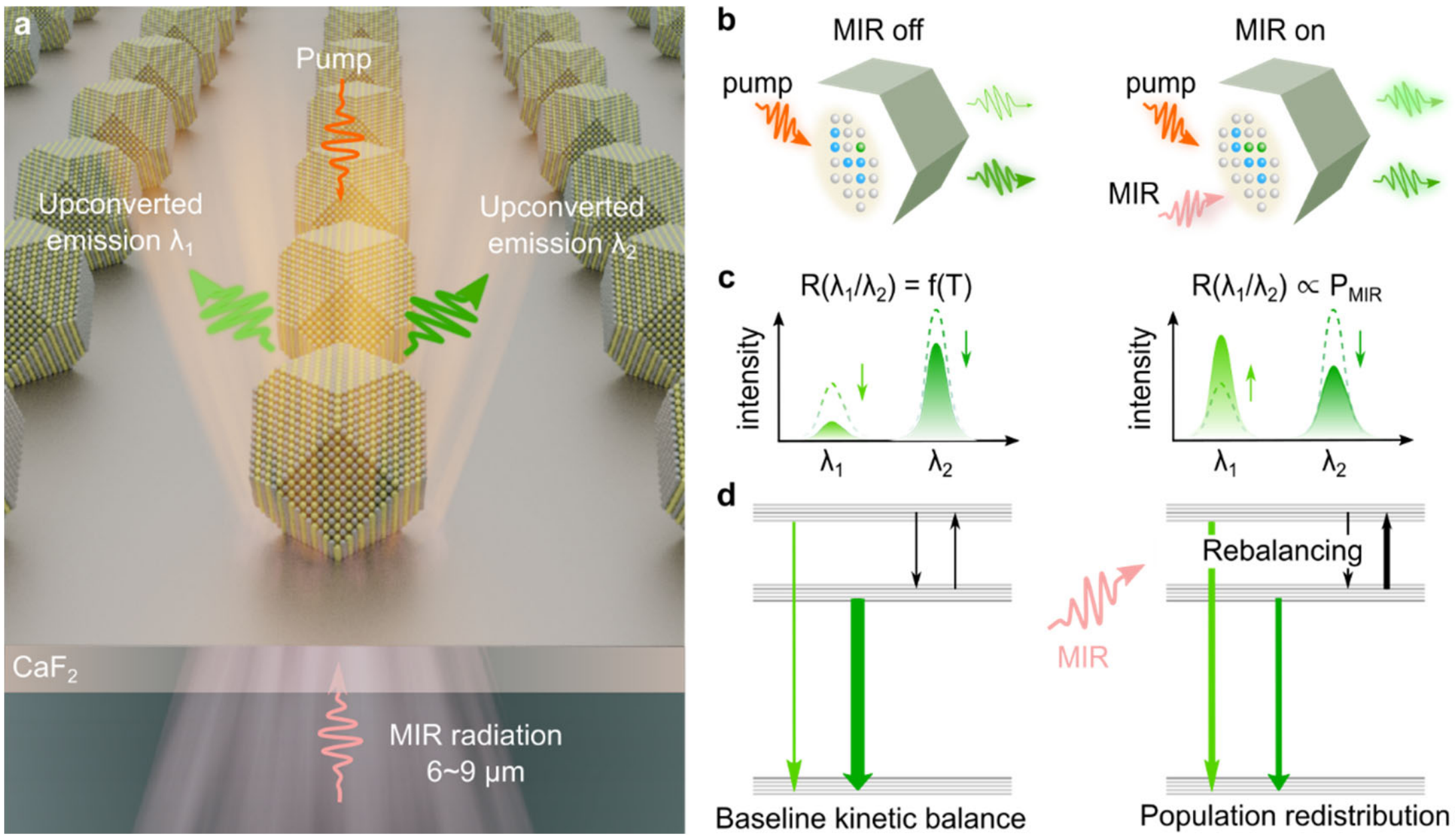

**Figure 1 │ Mid-infrared reconfiguration of excited-state population flow in lanthanide nanocrystals. a,** Conceptual illustration of MIR-modulated upconversion. A near-infrared pump excites lanthanide-doped nanocrystals on a $CaF_2$ substrate, generating two spectrally resolved visible emission channels ($\lambda_1$ and $\lambda_2$). Continuous-wave MIR radiation (6.8–8.6 μm) perturbs the excited-state relaxation network. **b,** Schematic comparison of the emission response with MIR off and on. MIR irradiation produces anticorrelated changes in the two emission channels, reflecting redistribution of their relative populations. **c,** Comparison between thermally driven and MIR-driven population redistribution. Thermal perturbation produces a temperature-dependent emission ratio, whereas MIR irradiation induces a distinct power-dependent ratiometric response. **d,** Schematic of the proposed kinetic reconfiguration in $Er^{3+}$. Population exchange between the thermally coupled manifolds is governed by competing relaxation pathways. MIR irradiation induces state-dependent changes in the excited-state kinetics, altering this kinetic balance and redistributing population between the emitting manifolds.

In thermally coupled lanthanide manifolds, phonon-assisted transitions mediate population exchange between closely spaced excited states. Under quasi-equilibrium conditions, the relative populations of these states are strongly influenced by lattice temperature, giving rise to the well-established Boltzmann dependence used in lanthanide luminescence thermometry. The MIR transduction scheme explored here is illustrated in **Fig. 1a**. $Yb^{3+}/Er^{3+}$ nanocrystals deposited on a $CaF_2$ substrate are excited by a near-infrared pump to generate spectrally resolved upconversion emission channels, while continuous-wave MIR radiation is applied simultaneously as an additional optical perturbation.

The MIR-induced redistribution of steady-state emission is illustrated schematically in **Fig. 1b**.[32] Without MIR irradiation, population exchange between the thermally coupled $Er^{3+}$ manifolds establishes a characteristic emission ratio determined by the underlying excitation and relaxation kinetics. Under MIR illumination, the relative populations of the excited states are rebalanced, leading to anticorrelated changes in the two emission channels. Consequently, the emission ratio exhibits a distinct MIR-power-dependent response compared with the predominantly temperature-dependent behaviour observed under thermal perturbation (**Fig. 1c**). As illustrated schematically in **Fig. 1d**, population exchange between the $^2H_{11/2}$ and $^4S_{3/2}$ manifolds is governed by competing upward and downward relaxation pathways. MIR irradiation perturbs the balance between these pathways, thereby altering the steady-state population distribution and the resulting emission ratio.

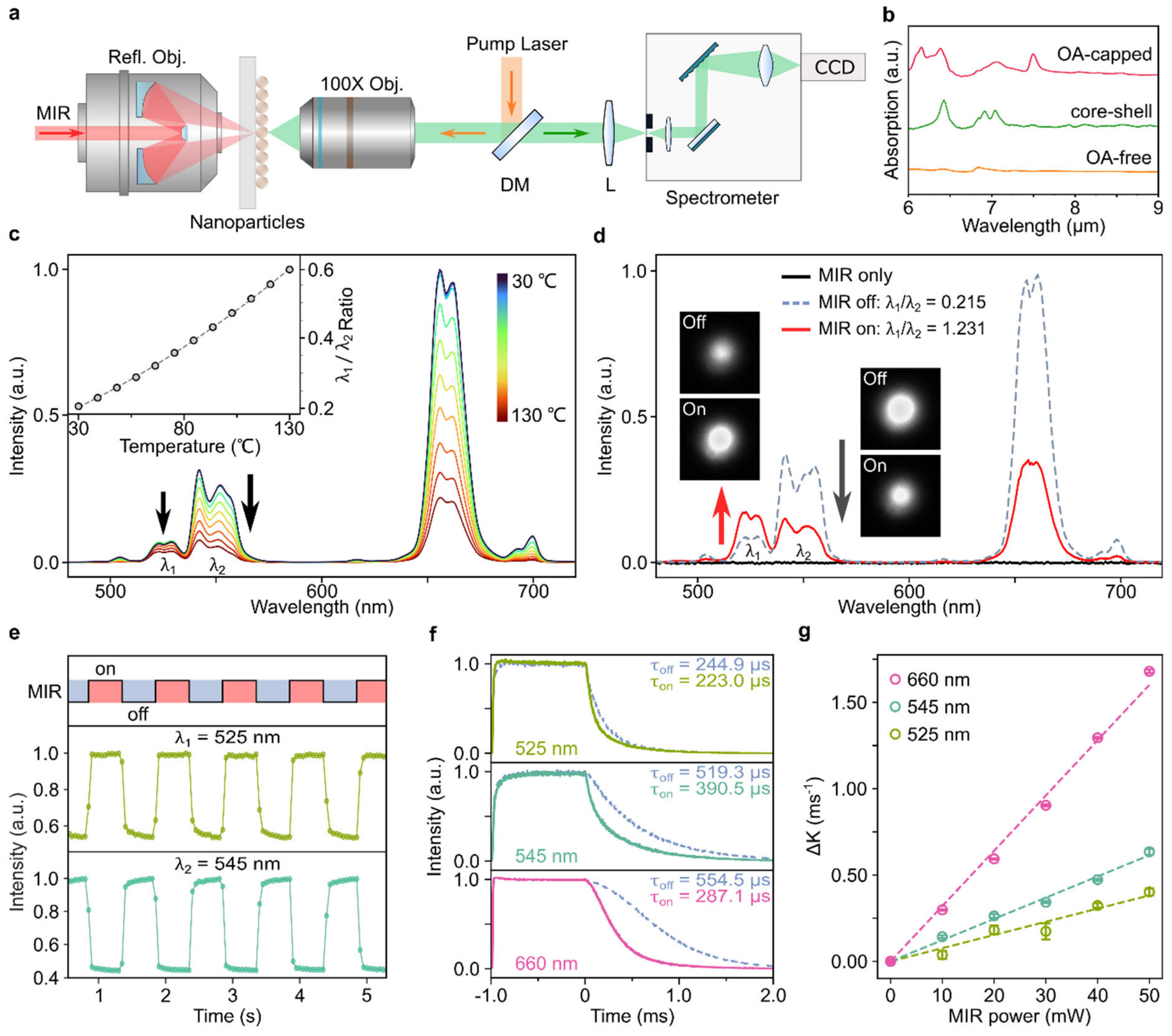

**Figure 2 │ State-dependent kinetics underlie MIR-induced population redistribution. a,** Schematic for MIR-modulated upconversion measurements. A tuneable MIR quantum cascade laser illuminates the nanocrystal ensemble while a 980 nm pump drives $Yb^{3+}/Er^{3+}$ upconversion. The converted visible emission is collected and analysed by the spectrometer. **b,** FTIR absorption spectra of OA-capped core, core-shell and OA-free nanocrystals. **c,** Thermal baseline measurement. Increasing temperature from 30 °C to 130 °C produces synchronous decrease of the 525 nm and 545 nm emission bands, while the intensity ratio changes from approximately 0.2 to 0.6. **d,** MIR-induced modulation under constant 980 nm pumping conditions. MIR illumination enhances the 525 nm emission while suppressing the 545 nm

band, producing a pronounced ratiometric contrast $I_{525}/I_{545}$: 0.22 → 1.23. Insets show corresponding upconversion images highlighting the MIR-induced ratiometric change. **e,** Time-domain reversible anticorrelated modulation of the 525 and 545 nm intensities under periodic MIR on/off illuminations. **f,** Time-resolved photoluminescence under 7.4 μm MIR irradiation at 50 mW. The $^2H_{11/2}$ (525 nm) lifetime changes marginally (~8.9%), whereas the $^4S_{3/2}$ (545 nm) and $^4F_{9/2}$ (660 nm) lifetimes shorten by ~24% and ~48.2%, respectively. **g,** MIR-power dependence of the change in inverse lifetime, $\Delta k(P) = \frac{1}{\tau(P)} - \frac{1}{\tau_0}$, for the three emission channels, showing approximately linear and strongly state-dependent kinetic responses.

To experimentally investigate MIR-induced reconfiguration of the excited-state kinetics, we built a dual-wavelength excitation platform, as illustrated in **Supplementary Note 1** and **Fig. S1**. Under 980 nm excitation, $Yb^{3+}$ sensitizers absorb pump photons and transfer energy to $Er^{3+}$ ions, populating excited states within the $Er^{3+}$ manifold that generate the characteristic green and red upconversion emission bands (**Supplementary Fig. S2**). A $NaYF_4$: 20%$Yb^{3+}$/2%$Er^{3+}$ nanocrystal film (see synthesis detail in **Supplementary Note 2**) was simultaneously illuminated by a 980 nm pump laser and a tuneable MIR quantum cascade laser as shown in **Fig. 2a**. The two beams were incident from opposite directions and spatially overlapped within the nanoparticle film, enabling controlled perturbation of the excited-state relaxation dynamics while maintaining constant optical excitation. Transmission electron microscopy confirms monodisperse nanocrystals with an average diameter of ~34 nm (**Supplementary Fig. S3**), suitable for ensemble characterization. Fourier-transform infrared spectroscopy reveals pronounced absorption within the MIR excitation range for the OA-capped nanocrystals (**Fig. 2b**). Following oleate removal, these absorption features are substantially suppressed, confirming effective removal of the MIR-active surface ligands.

To establish the response expected from conventional thermal perturbation, we first measured the temperature dependence of the upconversion emission. As shown in **Fig. 2c,** increasing the sample temperature from 30 °C to 130 °C leads to a simultaneous decrease of emission intensities across all bands (525 nm, 545 nm and 660 nm), due to the enhanced nonradiative relaxation. The intensity ratio between the two thermally coupled green emissions ($I_{525}/I_{545}$) increases gradually from 0.2 to 0.6 (**Fig. 2c** inset). The gradual increase in $I_{525}/I_{545}$ is consistent with thermally assisted population exchange between the $^4S_{3/2}$ and $^2H_{11/2}$ manifolds. In contrast, MIR illumination produces opposite modulation of the two thermally coupled emission bands. As illustrated in **Figure 2d**, the 525 nm emission increases while the 545 nm emission decreases, producing a ratio change from 0.22 to 1.23, substantially exceeding the ratio variation observed under direct thermal perturbation. Infrared thermal imaging under the same MIR excitation conditions shows a maximum sample temperature of approximately 62 °C (**Supplementary Figs. S4 and S5**). Direct heating to comparable temperatures reproduces neither the direction nor the magnitude of the MIR-induced spectral redistribution, indicating that uniform sample heating alone is insufficient to account for the measured response.

To further assess the contribution of ligand-mediated photothermal effects, we repeated the MIR measurements using OA-free nanocrystals. Despite the substantial suppression of oleate absorption shown in Fig. 2b, the ligand-free nanocrystals retain the characteristic anticorrelated

MIR response, with enhancement of the 525 nm emission and suppression of the 545 nm emission (**Supplementary Fig. S6**). These results show that oleate absorption is not required for the observed MIR-induced population redistribution, although surface ligands may influence the modulation magnitude.

The MIR-induced redistribution is reversible and dynamically correlated with the optical perturbation. Under periodic MIR on/off excitation, the two green bands respond reproducibly with opposite phases, as shown in **Fig. 2e**. Time-resolved photoluminescence measurements provide further evidence for state-dependent changes in the excited-state kinetics (**Fig. 2f**). The lifetime of the $^2H_{11/2}$ level (525 nm) changes only marginally from 244.9 to 223.0 μs (~8.9%), whereas the $^4S_{3/2}$ lifetime decreases from 519.3 to 390.5 μs (~24.8%). An even stronger reduction is observed for the $^4F_{9/2}$ level, from 554.5 to 287.1 μs (~48.2%). The markedly different responses of the three manifolds demonstrate that MIR irradiation does not simply produce a uniform rescaling of the emission dynamics but instead modifies the excited-state kinetics in a strongly state-dependent manner.

To quantify the MIR-induced kinetic response, we define the change in the apparent decay rate as

$$\Delta k(P) = \frac{1}{\tau(P)} - \frac{1}{\tau_0}$$

Where $k = 1/\tau$ denotes the apparent decay rate, $\tau(P)$ and $\tau_0$ denote the measured lifetimes at MIR power $P$ and with the MIR excitation switched off, respectively. As shown in **Fig. 2g** and **Supplementary Fig. S7**, $\Delta k$ increases approximately linearly with MIR power $P_{\mathrm{MIR}}$ over the measured range. Here. This power dependence can be described phenomenologically as:

$$\frac{1}{\tau(P)} = \frac{1}{\tau_0} + \alpha P_{MIR}$$

where $\alpha$ is an empirical coefficient that quantifies the sensitivity of the apparent decay kinetics to MIR power. The values of $\alpha$ are strongly state dependent, with the largest response observed for the $^4F_{9/2}$ manifold, followed by $^4S_{3/2}$, while $^2H_{11/2}$ is only weakly affected.

To identify the kinetic processes capable of reproducing the observed behaviour, we developed an expanded rate-equation framework incorporating the relevant $Er^{3+}$ manifolds, $Yb^{3+} \rightarrow Er^{3+}$ energy-transfer pathways and population-dependent relaxation processes (**Supplementary Note 3**). The experimentally observed anticorrelated emission and state-dependent lifetime changes can be reproduced by including an MIR-dependent perturbation of the excited-state kinetics (**Supplementary Fig. S8**). This model provides a phenomenological description consistent with MIR-driven reconfiguration of population flow within the coupled excited-state network.

**Excitation-decoupled ratiometric MIR transduction**

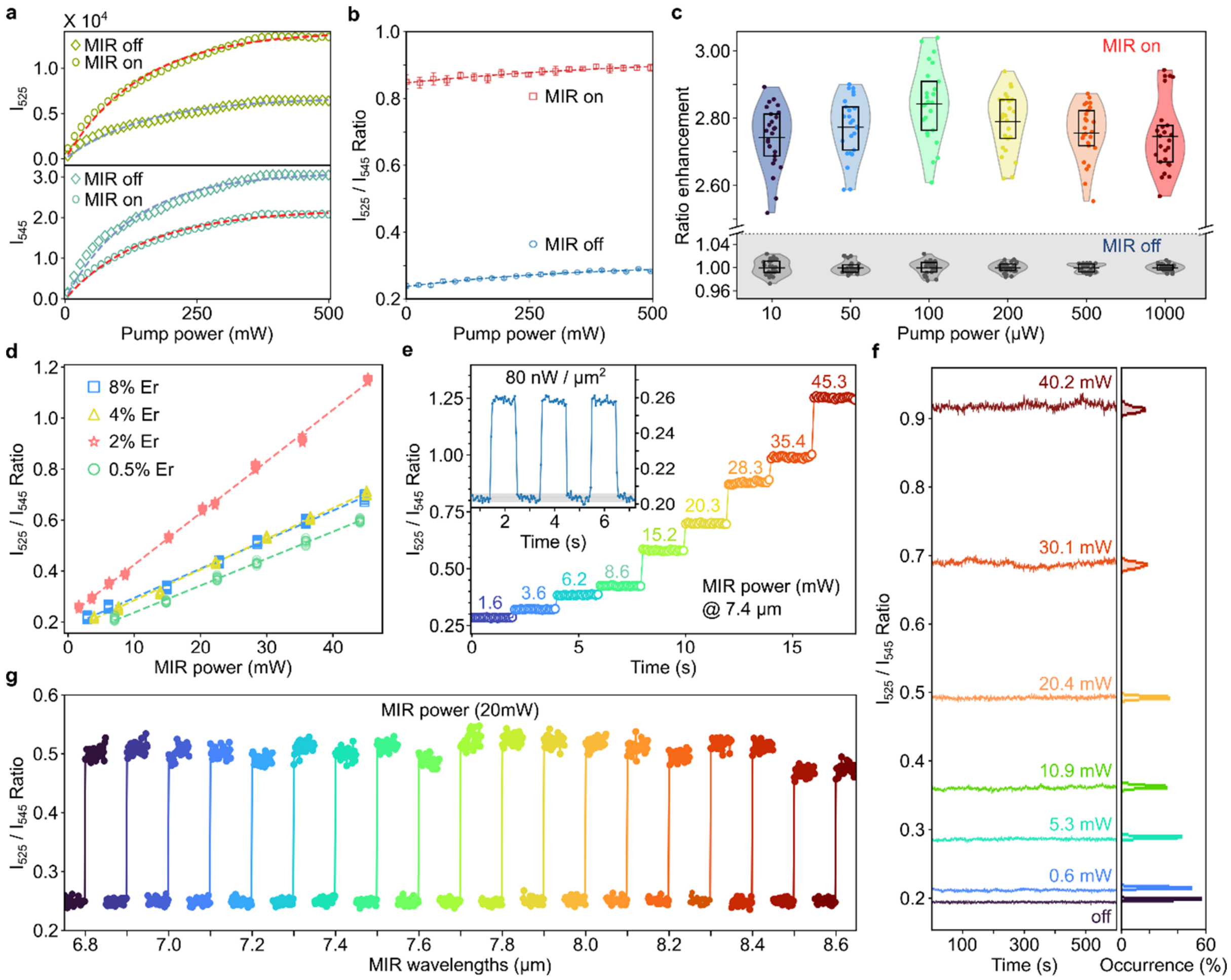

**Figure 3 │ Robust and excitation-decoupled ratiometric MIR transduction. a,** Pump-power dependence of the 525 nm and 545 nm emission intensities with and without MIR irradiation, showing opposite MIR-induced modulation across the excitation range. **b,** Emission ratio $I_{525}/I_{545}$ as a function of 980 nm pump power under MIR on and off conditions. **c,** Distribution of the ratiometric enhancement $R_{on}/R_{off}$ at pump powers from 10 to 1000 µW, demonstrating stable MIR contrast in the weak-excitation regime. **d,** MIR-power-dependent ratiometric response of nanocrystals with $Er^{3+}$ concentrations ranging from 0.5% to 8%. **e,** Stepwise ratiometric response to increasing MIR power at 7.4 µm under constant excitation. Inset, reversible modulation at a MIR power density of 80 nW µm$^{-2}$. The detection limit is approximately 4 nW µm$^{-2}$ at 7.4 µm, defined at a signal-to-noise ratio of 1 (**Supplementary Note 4**). **f,** Ratiometric signals measured at different MIR powers, with the corresponding occurrence distributions shown on the right. **g,** Reversible ratiometric response across the accessible MIR wavelength range of 6.8–8.6 µm.

To further examine how MIR irradiation modifies the excited-state dynamics, we investigated the pump-power dependence of the two green emission bands. In the absence of MIR irradiation, both the 525 nm ($^2H_{11/2}$) and 545 nm ($^4S_{3/2}$) emissions increase nonlinearly with pump power while remaining strongly correlated. As shown in **Fig. 3a**, this behaviour reflects conventional upconversion dynamics in which excitation density governs population generation through intermediate states. In contrast, MIR irradiation produces opposite modulation of the two thermally coupled bands across the entire pump-power range. The 525 nm emission is enhanced relative to the MIR-off state, whereas the 545 nm emission is suppressed. Consequently, MIR irradiation shifts the emission ratio to a distinct higher level. As shown in **Fig. 3b** and **Supplementary Figs. S9 and S10**, the ratiometric signal remains nearly constant over a broad excitation range from 10 μW to 500 mW.

Importantly, the overall pump dependence of the individual emission intensities is largely preserved, indicating that the MIR perturbation is superimposed on the underlying upconversion excitation dynamics. The ratiometric response therefore remains largely insensitive to excitation density over three orders of magnitude of pump power (10–1000 μW, **Fig. 3c**). This separation between population generation by the near-infrared pump and MIR-induced population redistribution enables the ratiometric readout to remain largely decoupled from the excitation intensity.

The robustness of the MIR-induced ratiometric response was further examined across $Er^{3+}$ concentrations ranging from 0.5% to 8%, as shown in **Fig. 3d**. Although the absolute emission ratios vary with $Er^{3+}$ concentration, reflecting changes in the underlying upconversion and energy-transfer dynamics, all samples retain an approximately linear dependence of $I_{525}/I_{545}$ on MIR power. This indicates that the MIR-induced population redistribution is not restricted to a narrowly defined doping condition. Bare-core and core–shell nanocrystals also exhibit MIR-dependent ratiometric modulation (**Supplementary Fig. S11**), although the core–shell particles show a reduced modulation amplitude. This structural dependence indicates that the MIR response is sensitive to changes in the local excited-state relaxation environment, consistent with a kinetic origin of the population redistribution.

At fixed 980 nm excitation, the emission ratio increases approximately linearly with MIR power (**Fig. 3e**). Stepwise changes in MIR power produce clearly resolved ratiometric levels, while periodic MIR modulation results in reversible transitions between the corresponding ratiometric levels. The approximately linear calibration provides a detection limit of ~4 nW $\mu m^{-2}$ at 7.4 μm, defined at a signal-to-noise ratio of 1 (**Supplementary Note 4**). The statistical distributions shown in **Fig. 3f** further demonstrate progressive separation of the ratiometric signal with increasing MIR power, supporting quantitative discrimination between different MIR excitation levels.

To further examine the spectral range of the response, we measured the ratiometric modulation across the accessible QCL tuning range. As shown in **Fig. 3g**, reversible MIR-induced modulation persists from 6.8 to 8.6 μm. At each wavelength, the emission ratio varies approximately linearly with MIR power (**Supplementary Fig. S12**), demonstrating broadband and quantitative MIR transduction within the measured spectral window. The wavelength dependence is smooth across this range and shows no pronounced narrow resonant feature. Because the accessible range is limited by the QCL source, these measurements define the experimentally demonstrated bandwidth rather than the full intrinsic response spectrum of the nanocrystals. The nanocrystal response is also photostable under prolonged dual-wavelength excitation. Under continuous 980 nm excitation with MIR co-irradiation, the emission ratio measured from a fixed region remains stable for more than 5 h, with a standard deviation of 3.74%, comparable to the <3% power fluctuation of the QCL source (**Supplementary Fig. S13**).

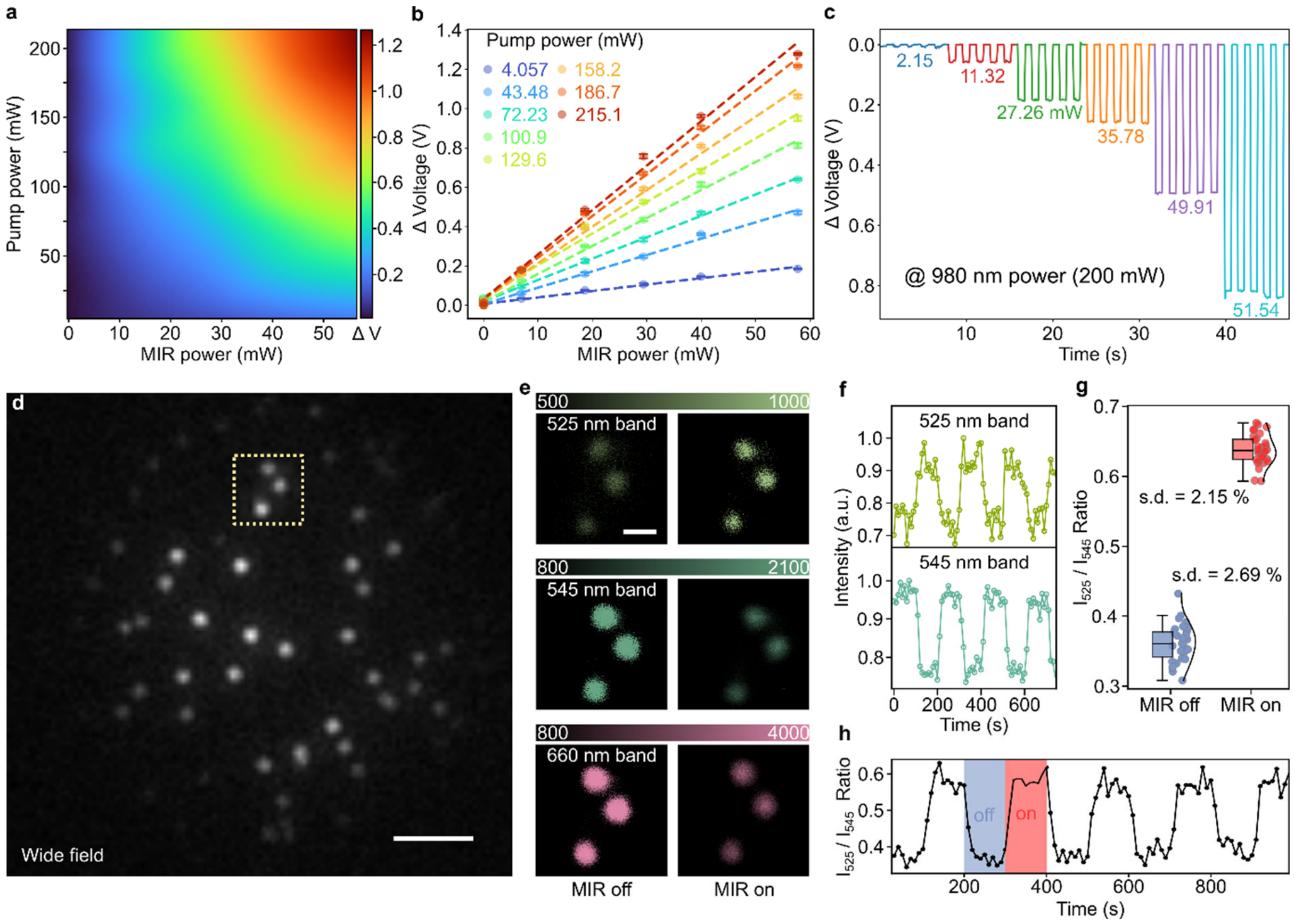

**Figure 4 | Silicon-compatible MIR detection and single- nanocrystal ratiometric imaging. a,** Electrical response of the lanthanide nanocrystal film integrated with a silicon photodetector. The colour map shows the measured voltage change ΔV as a function of MIR power and 980 nm pump power under 660 nm emission detection. **b,** Linear dependence of $\Delta V$ on MIR power at different 980 nm pump powers **c,** Time-resolved electrical response under stepwise MIR excitation, showing stable and reproducible modulation. **d,** Wide-field fluorescence image of sparsely dispersed nanocrystals used for single- nanocrystal measurements. The dashed region highlights representative emitters. Scale bars, 5 μm. **e,** Representative single- nanocrystal emission images in the 525, 545 and 660 nm channels with MIR off and on. Scale bars, 1 μm. **f,** Time-resolved emission intensities of the 525 and 545 nm channels from a representative single nanocrystal under periodic MIR excitation. **g,** Statistical distribution of the single-nanocrystal ratiometric response $I_{525}/I_{545}$with and without MIR illumination. **h,** Reversible ratiometric modulation from an individual nanocrystal under repeated MIR on/off excitation.

To evaluate compatibility with conventional optical detectors, we coupled the lanthanide nanocrystal film to a silicon photodetector for electrical readout of the MIR-induced upconversion modulation. The 660 nm emission, which exhibits the largest MIR-induced intensity change, was spectrally isolated using a band-pass filter and detected as a voltage change $\Delta V$. For these measurements, the MIR beam was mechanically chopped at 100 Hz, and the photodetector signal was recovered using lock-in detection (**Supplementary Fig. S1**). As shown in **Fig. 4a**, $\Delta V$ increases systematically with both MIR and pump power, demonstrating conversion of the MIR perturbation into an electrically measurable visible signal.

The electrical response varies approximately linearly with MIR power over the measured range at different 980 nm excitation powers (**Fig. 4b** and **Supplementary Fig. S14**). Stepwise changes in MIR power produce stable and reproducible electrical modulation (**Fig. 4c**), demonstrating quantitative MIR detection using a standard silicon photodetector. In the present measurements, the modulation frequency was fixed at 100 Hz by the mechanical chopping and lock-in detection scheme; therefore, the measured temporal response reflects the instrumental operating condition rather than the intrinsic bandwidth of the nanocrystal transducer.

Comparison with representative MIR upconversion platforms highlights the low excitation requirements of the present system. The nanocrystal transducer operates at room temperature and maintains measurable ratiometric response under near-infrared excitation powers down to the microwatt regime, while retaining compatibility with silicon photodetectors and conventional visible-light optics. A broader comparison of operating wavelengths, excitation conditions and detection performance is provided in **Supplementary Table S1 and Fig. S15**.

We next examined whether the MIR-induced kinetic redistribution persists at the single-nanocrystal level. Wide-field fluorescence imaging was performed on sparsely dispersed nanocrystals (**Fig. 4d** and **Supplementary Fig. S16**). Representative single-nanocrystal images acquired in the 525, 545 and 660 nm spectral channels are shown in **Fig. 4e**. Under MIR illumination, the 525 nm emission increases whereas the 545 and 660 nm emissions decrease, reproducing the characteristic spectral redistribution observed in the ensemble measurements.

Time-resolved measurements from individual nanocrystals further reveal reversible and anticorrelated modulation of the 525 and 545 nm emission channels under periodic MIR excitation (**Fig. 4f**). The corresponding ratio $I_{525}/I_{545}$ shifts reproducibly to higher values under MIR illumination across the measured particles, as shown in **Fig. 4g**. Repeated MIR on/off modulation of an individual nanocrystal produces reversible changes of the ratiometric signal without detectable degradation over the measurement period (**Fig. 4h**). These measurements demonstrate that the MIR-induced redistribution persists at the single-nanocrystal level and is not a consequence of ensemble averaging.

## Discussion

The results presented here show that incoherent MIR irradiation can reconfigure population flow within the coupled excited-state network of $Yb^{3+}/Er^{3+}$ nanocrystals. In thermally coupled lanthanide manifolds, population exchange is strongly influenced by phonon-assisted relaxation and lattice temperature.[33,34] Here, MIR illumination produces anticorrelated emission changes together with strongly state-dependent lifetime modifications that cannot be reproduced by uniform thermal perturbation alone. The persistence of the characteristic response after oleate removal further shows that absorption by surface oleate ligands is not required for the effect. Together, these observations support an MIR-driven redistribution of population among competing excited-state pathways.

This behaviour complements previously reported MIR–lanthanide transduction mechanisms based on excited-state absorption, ligand-mediated photothermal effects and temperature-dependent energy transfer.[27-31] In the present $Yb^{3+}/Er^{3+}$ system, the combination of anticorrelated spectral redistribution, state-dependent changes in the apparent decay rates and ligand-independent MIR response points to a distinct kinetic regime in which MIR irradiation modifies how excitation is redistributed among coupled $Er^{3+}$ manifolds.[35] The near-infrared pump establishes the excited-state population, whereas MIR irradiation perturbs its subsequent redistribution, allowing the ratiometric output to remain largely decoupled from excitation intensity.

This population-flow reconfiguration enables linear ratiometric MIR transduction across 6.8–8.6 μm, operation at near-infrared pump powers down to 10 μW and a detection limit of approximately 4 nW $\mu m^{-2}$ using standard silicon photodetectors. Importantly, the characteristic anticorrelated response persists at the single-nanocrystal level, demonstrating that the effect is not a consequence of ensemble averaging. More broadly, these results position lanthanide nanocrystals not only as spectral converters, but as optically reconfigurable nanoscale energy-flow networks in which low-energy infrared radiation reshapes the distribution of excitation among competing emissive states. Such reconfiguration provides a route towards low-power MIR transduction and integration with silicon-compatible and computational photonic architectures.[36,37]

## Online Methods

All the experimental details, including the synthesis of nanocrystals, structural characterization, optical set-up, data acquisition and numerical simulations of the MIR modulation process, are provided in the Supplementary Information.

## Data availability

All relevant data that support the findings of this work are available from the corresponding author on reasonable request.

## Code availability

All code used in this paper is available from the corresponding authors upon reasonable request.

## Acknowledgements

The authors acknowledge financial support from the Australian Research Council (DE250100406, DP260104454, CE200100010, FT220100053).

## Author contributions

C.C. and I.A. designed and conceptualized the research. C.C. and I.A. supervised the work. X.Y., K.Y., and C.C. built the mid-infrared optical setup. X.Y., C.C., M.K. and J.K. performed optical experiments and data analysis. Y.H., G.B. and X.X. synthesized the nanoparticles and performed the electron microscopy imaging. C.W., C.C. and F.W. performed the numerical simulations. L.D., B.L. and O.S. fabricated the nanoscale test chart. X.Y., Y.Y., Y.B., L.F., Y.H. and Q.S. performed the FTIR and temperature control measurements. X.Y., C.C. and I.A. wrote the manuscript. All authors discussed the results and revised the manuscript.

## Competing interests

The authors declare no competing interests.